\documentclass[floatfix,aps,prl,notitlepage,amsmath,amssymb,superscriptaddress,10pt,nobibnotes,twocolumn]{revtex4-1}
\pdfoutput=1

\usepackage{amsfonts, amssymb, amsmath, dsfont,mathrsfs}
\usepackage{graphicx}
\usepackage{color}
\usepackage[utf8]{inputenc}
\usepackage[T1]{fontenc}
\usepackage[svgnames]{xcolor}
\usepackage{textcomp}
\usepackage{siunitx}
\usepackage{multibib}
\newcites{supp}{Supplementary References}

\usepackage{hyperref}
\hypersetup{colorlinks,urlcolor=DarkBlue,linkcolor=DarkBlue,citecolor=DarkBlue,pdfdisplaydoctitle=true,pdfpagemode=UseOutlines,bookmarksnumbered=true,bookmarksopen=true} 

\makeatletter
\renewcommand{\eqref}[1]{\textup{\tagform@{\ref*{#1}}}}
\makeatother

\makeatletter
\renewcommand{\figurename}{Fig.}
\renewcommand*{\fnum@figure}{{\normalfont\bfseries \figurename~\thefigure}}
\makeatother

\renewcommand{\bm}[1]{\boldsymbol{\mathbf{#1}}} 

\newcommand{\beginsupplement}{%
        \setcounter{table}{0}
        \renewcommand{\thetable}{S\arabic{table}}%
        \setcounter{figure}{0}
        \renewcommand{\thefigure}{S\arabic{figure}}%
        \setcounter{section}{0}
        \renewcommand{\thesection}{\Roman{section}}
        \setcounter{equation}{0}
        \renewcommand{\theequation}{S\arabic{equation}}
     }

\begin{document}

\title{Maximum-likelihood estimation in ptychography in the presence of Poisson-Gaussian noise statistics}

\author{Jacob Seifert}
\altaffiliation{\href{mailto:j.seifert@uu.nl}{j.seifert@uu.nl}}
\affiliation{Nanophotonics, Debye Institute for Nanomaterials Science and Centre for Extreme Matter and Emergent Phenomena, Utrecht University, P.O. Box 80000, 3508 TA Utrecht, The Netherlands}
\author{Yifeng Shao}
\affiliation{Imaging Physics Department, Applied Science Faculty, Delft University of Technology, The Netherlands}
\author{Rens van Dam}
\affiliation{Nanophotonics, Debye Institute for Nanomaterials Science and Centre for Extreme Matter and Emergent Phenomena, Utrecht University, P.O. Box 80000, 3508 TA Utrecht, The Netherlands}
\author{Dorian Bouchet}
\affiliation{Université Grenoble Alpes, CNRS, LIPhy, 38000 Grenoble, France}
\author{Tristan van Leeuwen}
\affiliation{Mathematical Institute, Utrecht University, Budapestlaan 6, 3584CD, Utrecht, The Netherlands}
\affiliation{Centrum Wiskunde \& Informatica, Science Park 123, 1098 XG, Amsterdam, The Netherlands}
\author{Allard P.\ Mosk}
\affiliation{Nanophotonics, Debye Institute for Nanomaterials Science and Centre for Extreme Matter and Emergent Phenomena, Utrecht University, P.O. Box 80000, 3508 TA Utrecht, The Netherlands}

\begin{abstract}
\noindent Optical measurements often exhibit mixed Poisson-Gaussian noise statistics, which hampers image quality, particularly under low signal-to-noise ratio (SNR) conditions. Computational imaging falls short in such situations when solely Poissonian noise statistics are assumed. In response to this challenge, we define a loss function that explicitly incorporates this mixed noise nature. By using maximum-likelihood estimation, we devise a practical method to account for camera readout noise in gradient-based ptychography optimization.
Our results, based on both experimental and numerical data, demonstrate that this approach outperforms the conventional one, enabling enhanced image reconstruction quality under challenging noise conditions through a straightforward methodological adjustment.
\end{abstract}

\maketitle


In the rapidly evolving field of computational imaging, ptychography has emerged as a powerful technique capable of producing high-resolution phase and amplitude images from diffraction patterns. 
It involves translating a thin object through overlapping illuminations and measuring the resulting diffraction patterns behind the object with a camera sensor~\cite{Rodenburg2004-oi, Rodenburg2019-kw}. Subsequently, the complex-valued image is constructed through an iterative optimization algorithm, necessitating the formulation and minimization of a loss function, alternatively referred to as the objective, cost, or error function. 
Ptychography has found applications in a wide range of topics, including label-free biological imaging \cite{Giewekemeyer2010-at, Marrison2013-wz, Polo2020-cg}, optical metrology~\cite{Moore2016-ag, Claus2013-wv, Song2019-yo, Du2023-ib, Bouchet2021-mb}, and atomic-resolution imaging using electron beams~\cite{Yang2016-np, Kharitonov2021-yt, Chen2020-we}.

\begin{figure}[ht]
\centering
\includegraphics[width=0.9\linewidth]{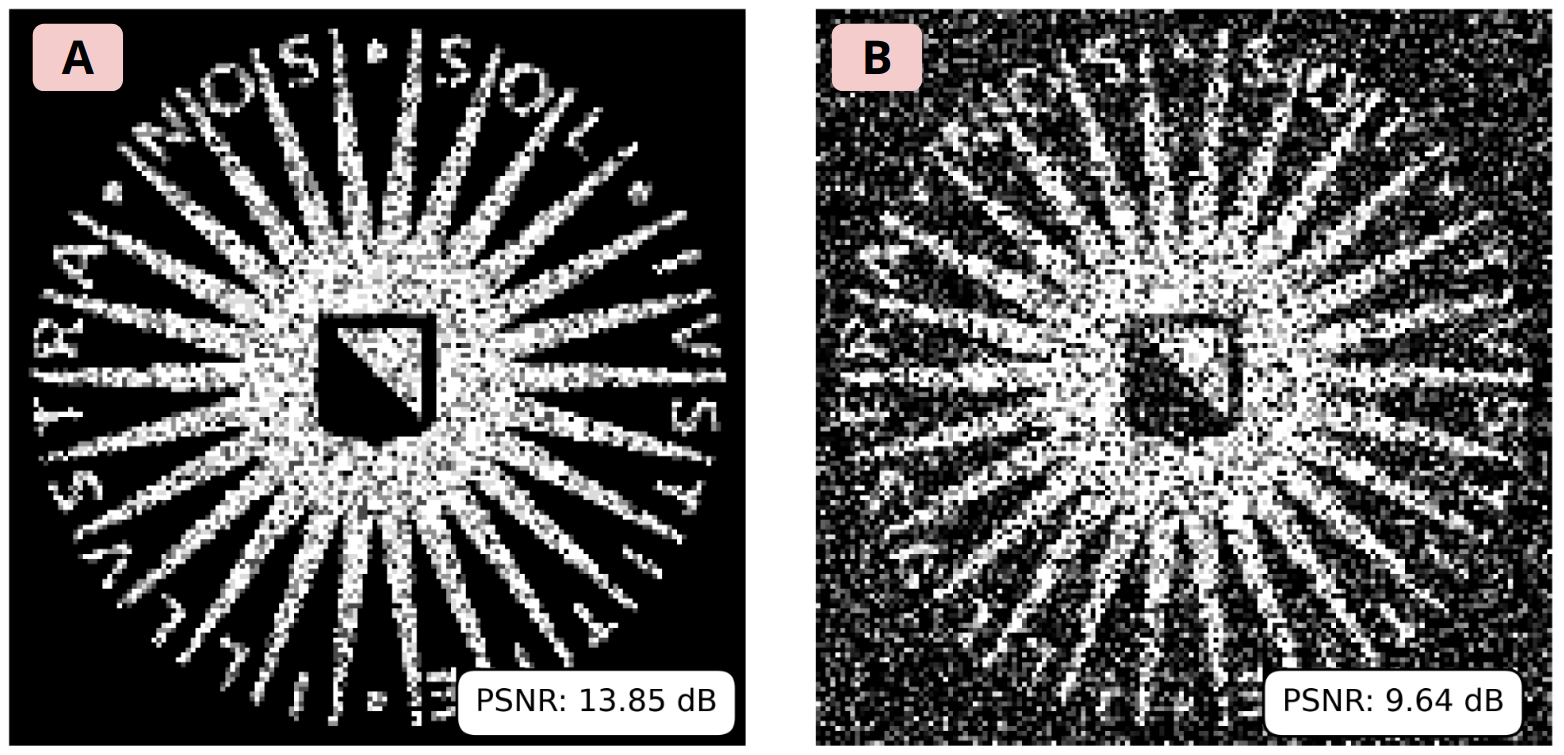}
\caption{Comparative visualization of a greyscale image distorted by different noise types. Panel (A) depicts an image with simulated Poissonian noise, while panel (B) illustrates the effect of simulated mixed Poisson-Gaussian noise resulting from additive readout noise. Inlaid values indicate Peak Signal-to-Noise Ratio (PSNR) with respect to the ground truth.}
\label{fig:noise}
\end{figure}

Fundamentally, the basis of ptychographic reconstructions is the detection of photon counts on a camera sensor and, therefore, subject to Poissonian noise even under ideal measurement conditions. Given the assumption of an underlying noise model, a powerful and robust optimization strategy is the maximum-likelihood estimation (MLE) principle~\cite{Kay1993-vj}. 
By leveraging MLE in ptychography, one seeks to estimate the studied object parameters that render the observed diffraction patterns most probable~\cite{Thibault2012-mn, Godard2012-gr, yeh2015experimental, Zhang2017-jr, Wei2020-wp}. 
However, additive camera readout noise is often neglected, thus leaving a gap in the fidelity of the reconstructions. This is a salient concern as the presence of readout noise, typically Gaussian, is an important element of practical ptychographic measurements within the visible spectrum. Ignoring this noise source oversimplifies the underlying statistical model and introduces errors to the reconstructed image, especially when the detected photon counts and signal-to-noise ratio (SNR) are low. The distinction between a Poissonian and a mixed Poisson-Gaussian noise model is depicted in the simulated images presented in Fig.~\ref{fig:noise}.

In this letter, we propose a loss function for automatic differentiation ptychography that explicitly incorporates both Poissonian and Gaussian noise sources. This approach brings us closer to the real-world conditions of ptychographic measurements, thereby paving the way for superior performance in image reconstruction under challenging noise conditions. 
We outline a practical method to incorporate camera readout noise in computational imaging. Furthermore, we provide a comprehensive comparison between the image reconstruction quality using a mixed-statistics loss function and that of a conventional loss function which presumes solely Poissonian noise statistics. For this, we present reconstruction results obtained from both experimental and numerical data.

\begin{figure}[t]
\centering
\includegraphics[width=0.45\textwidth]{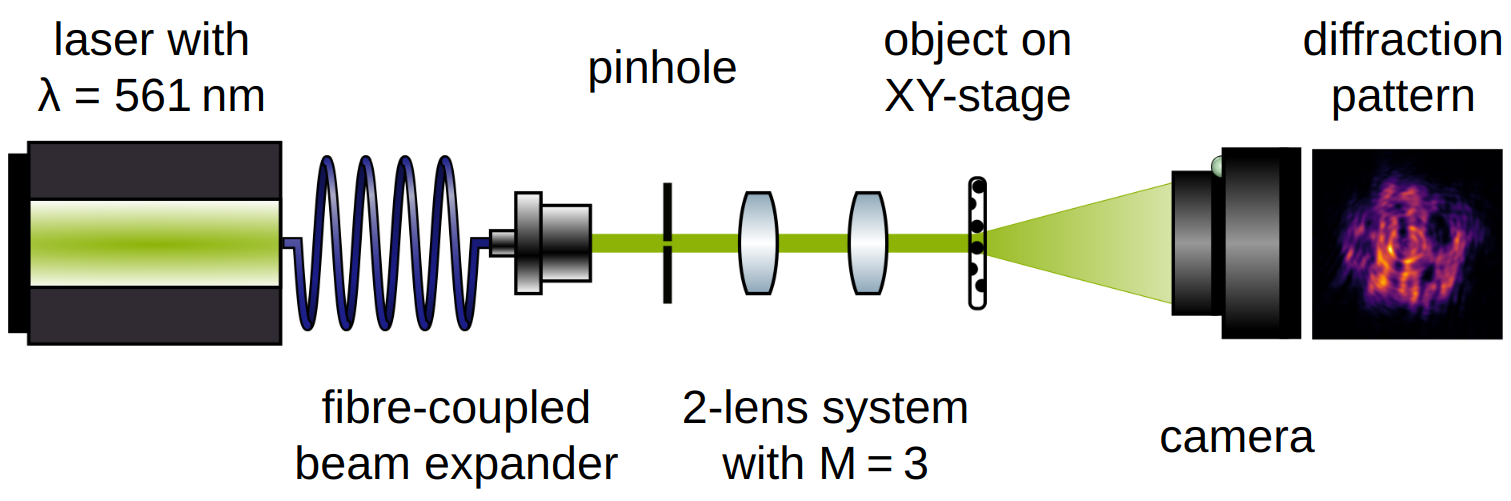}
\caption{Schematic drawing of our ptychography setup used in the experiment and for numerical simulations. A 500-µm pinhole is illuminated and relayed onto the object using a 2-lens system. The object laterally shifted through the beam using an XY-stage. A CMOS camera records the diffraction intensities at a distance of \SI{38}{mm} downstream of the object.}
\label{fig:setup}
\end{figure}

In ptychography, the typical reconstruction approach involves minimizing a loss function representing the difference between the intensity values of the observed diffraction pattern $X_k$ and the anticipated diffraction patterns $I_k(\bm{\theta})$ as determined by a parameter set $\bm{\theta}$, which embodies the object under investigation, at all pixel locations indexed by $k$.
In the presence of measurement noise, it is insightful to tackle the problem of ptychographic reconstruction by maximizing the likelihood of the observed given the object parameters. From this probabilistic perspective, one seeks the object parameters that make the observed data most likely, which renders object retrieval more robust in scenarios of low SNR. 

In the supplement (section 1), we elaborate on deriving the maximum-likelihood estimation (MLE) loss functions for two different types of noise statistics. When operating under the assumption of Poissonian counting noise, the loss function $L_{\mathrm{Poisson}}$ that yields the maximum-likelihood estimate is expressed as:
\begin{equation} \label{poisson_loss}
    L_{\mathrm{Poisson}}(\bm{\theta})=\sum_{k=1}^N\left( \sqrt{X_k}- \sqrt{I_k(\bm{\theta})}\right)^2,
\end{equation}
where the sum encompasses $N$ statistically independent pixels on the camera sensor. To account for Gaussian readout noise on the camera sensor, an additional data acquisition step becomes essential to extend the MLE loss function: the variance $\sigma_k^2$ of the readout noise of the camera must be determined using multiple full-frame dark images. With this additional information and the assumption that the Poissonian component of the statistics can be approximated by a Gaussian distribution, we formulate the MLE loss function that incorporates mixed Poisson-Gaussian noise statistics:
\begin{equation} \label{mixed_loss}
    L_{\mathrm{Mixed}}(\bm{\theta})=\sum_{k=1}^N\left( \ln[I_k(\bm{\theta}) + \sigma_k^2] + 
    \frac{[X_k- I_k(\bm{\theta})]^2}{I_k(\bm{\theta}) + \sigma_k^2} \right).
\end{equation}
Note that this expression is not only relevant for pixels with high photon counts where the Gaussian approximation of Poisson statistics is most accurate, but also for pixels with low photon counts. Indeed, for those low-count pixels, the Gaussian readout noise is the dominant source of noise such that the deviation of Poisson statistics from a Gaussian distribution becomes irrelevant.

\begin{figure*}[ht]
\centering
\includegraphics[width=0.68\linewidth]{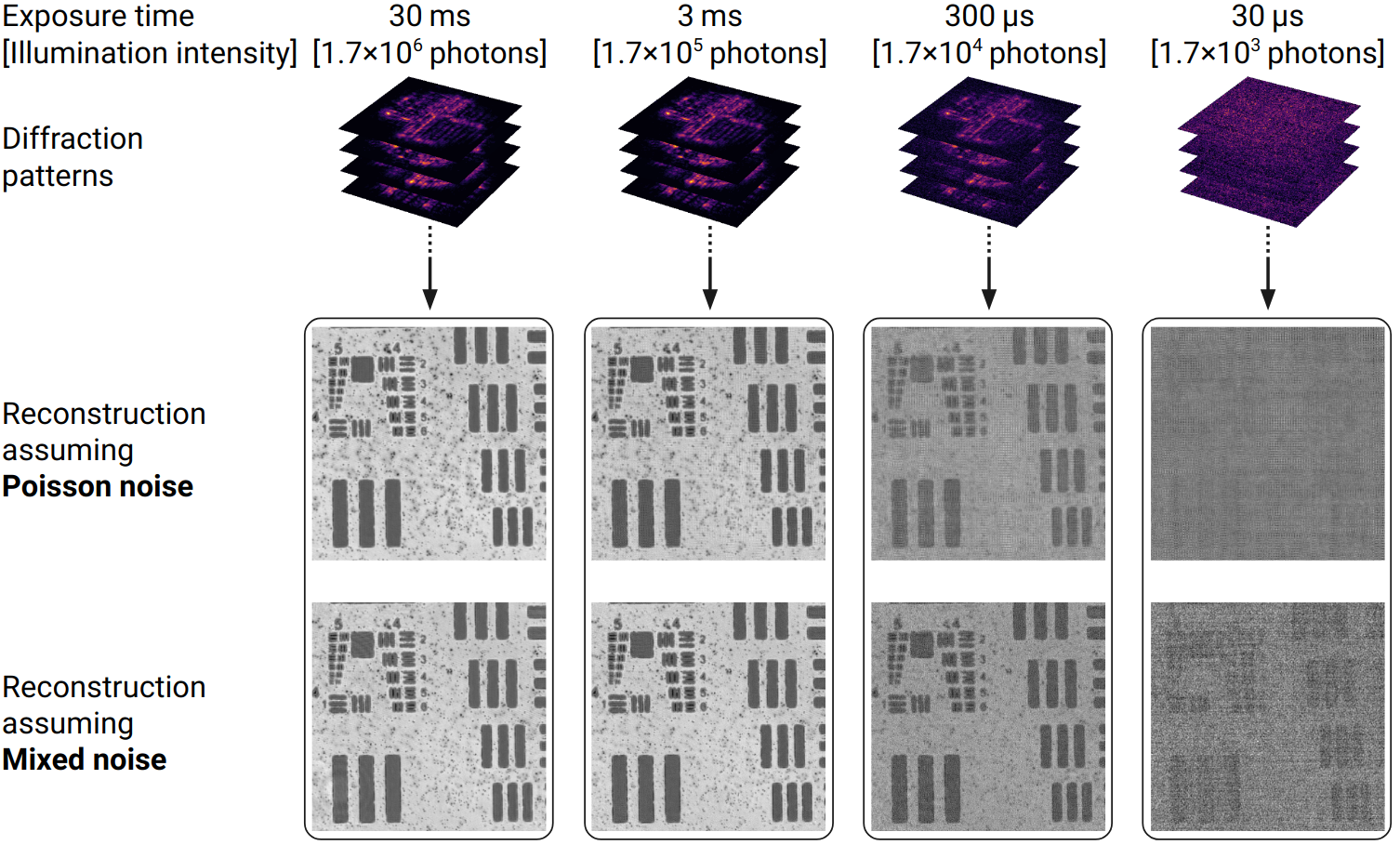}
\caption{Comparison of image reconstruction qualities from ptychographic datasets with decreasing signal-to-noise ratio (SNR) from left to right. Top row: Ptychographic datasets at different camera exposures with total illumination intensity, respectively. Second row: Reconstructions using a loss function assuming solely Poissonian noise statistics. Third row: Reconstructions using the loss function defined by Eq. \ref{mixed_loss} which incorporates mixed Poisson-Gaussian noise statistics. Edge length of every image equals \SI{3.5}{mm}.}
\label{fig:experiment}
\end{figure*}

To validate the beneficial effect of the loss function $L_{\mathrm{Mixed}}(\bm{\theta})$ experimentally, we are considering a standard ptychography setup in a transmission geometry (Fig.~\ref{fig:setup}). A circular 500-µm pinhole is illuminated with coherent light with wavelength \SI{561}{nm} and relayed to the sample plane using two lenses with a magnification $M=3$. 
There, a binary target sample is illuminated at 80 scanning positions with an overlap of approximately \SI{60}{\percent} between adjacent positions, which lies within the ideal regime according to \cite{Bunk2008-nq}. 
The scattered light is captured by a CMOS camera positioned \SI{38}{mm} away from the object. To ensure a comprehensive comparison as a function of SNR, we employ four different exposure settings per scanning position, spanning a range from \SI{30}{\micro\second} to \SI{300}{\milli\second}, with each subsequent exposure time differing by a factor of 10 (see Fig.~\ref{fig:experiment}). For each exposure time, we determine the variances $\sigma_k^2$ by capturing a stack of 300 dark images. This quantifies the readout noise level associated with each pixel $k$ of our camera.
A comprehensive overview of further details about the experimental implementation and methodology can be found in section 2 of the supplementary information.


By maintaining a constant illumination power, we acquire four distinct ptychographic datasets, each corresponding to a different signal-to-noise ratio (SNR), as demonstrated in the top row of Fig.~\ref{fig:experiment}. With these datasets in hand, we proceed to perform image reconstructions utilizing two different loss functions (Eq. \ref{poisson_loss} and \ref{mixed_loss}) within an automatic differentiation-based ptychography framework, as detailed in \cite{Seifert2021-fr} and similar to \cite{Du2020-sh}. For this analysis, we precalibrate the illumination field using an additional high-SNR measurement and restrict our optimizations to the complex-valued object transmission functions, particularly under decreasing SNR conditions. This approach enables us to isolate the impact of the loss function choice from the convergence behavior associated with an unknown illumination field. The reconstruction procedure is explained in more detail in section 3 of the supplementary information, and the source code and raw data are available in \cite{Seifert2023-ak} under open licenses.
In scenarios characterized by high SNR, no noticeable disparity in image quality is observed between the two approaches. However, when confronted with low SNR conditions, where the signal becomes immersed within the readout noise, the advantages of the mixed-statistics MLE loss function become clear. The reconstructed images exhibit superior quality and reveal finer details that would otherwise remain obscured without accounting for the readout noise statistics.

For a quantitative analysis and validation of our experimental findings, we generate a simulated object with phase and amplitude contrasts which we treat as the ground truth. Using numerical simulation and the ground truth object, we compute the noisy diffraction patterns assuming Poissonian photon count statistics and an additive Gaussian readout noise with a standard deviation of $\sigma = \SI{1.5}{counts}$. The simulation allows for varying the illumination intensity given as the total number of photons in an otherwise fixed illumination field that approximates the experimental conditions shown in Fig.~\ref{fig:experiment}, thereby controlling the measurement SNR.

\begin{figure*}[ht]
\centering
\includegraphics[width=0.685\linewidth]{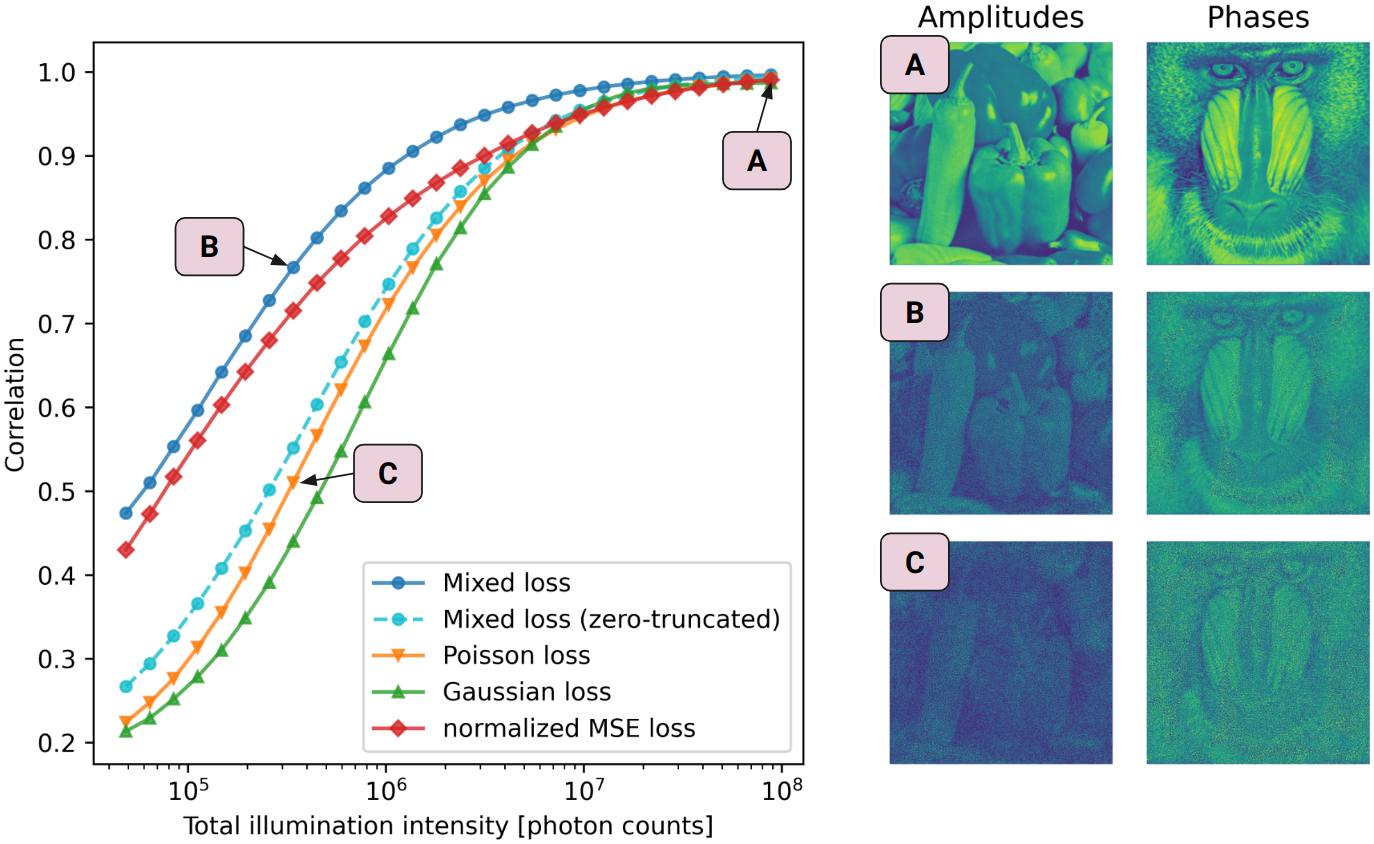}
\caption{Correlation between ground truth and reconstructions as a function of total photon count in the illumination field, based on numerical simulations. The dashed blue line represents reconstructions derived from intensity data where negative values have been zero-cropped, while the solid line represents reconstructions that incorporate negative values, which can arise due to the additive Gaussian noise component. (A) Amplitude and phase contrast reconstruction of the simulated object based on high-intensity diffraction patterns. For comparison, with an illumination intensity of \num{3.4d5}\,photons the reconstructions achieved with $L_\mathrm{Mixed}$ (B) and $L_\mathrm{Poisson}$ (C) are shown. The edge length of every image equals \SI{3.5}{mm}. }
\label{fig:simulation}
\end{figure*}

The achieved reconstruction quality of the object $O$ with respect to the ground truth $O_\mathrm{gt}$ can now be quantified using the correlation coefficient 
$C = \frac{{\left| \left\langle \overline{O}_\mathrm{gt}, O \right\rangle \right|}}{{\|O_\mathrm{gt}\| \cdot \|O\|}}$
as defined and motivated in \cite{Seifert2021-fr} as a function of illumination intensity (Fig.~\ref{fig:simulation}). Here, $\overline{O}_\mathrm{gt}$ denotes the complex conjugate of $O_\mathrm{gt}$, $\left\langle \cdot, \cdot \right\rangle$ denotes the dot product, and $\|\cdot\|$ denotes the norm.
The simulation confirms the same trend that we observe from the experiments: For low SNR, optimization using a mixed-statistics loss function yields significantly better reconstruction results. In the regime of high illumination intensities, all loss functions converge excellently (up to machine precision) as the readout noise becomes irrelevant, and all assumed underlying probability density functions become valid approximations.
Image reconstruction using a Gaussian loss function $L_{\mathrm{Gaussian}}(\bm{\theta})=\sum_{k=1}^N\left( X_k- I_k(\bm{\theta})\right)^2$ performs worst at low illumination intensities. 
However, introducing a weighting term as motivated in~\cite{mildenhall2021nerf} leads to a noteworthy improvement (shown in red) using $L_{\mathrm{normMSE}}(\bm{\theta})=\sum_{k=1}^N\left( \frac{X_k- I_k(\bm{\theta})}{\mathrm{sg}[I_k(\bm{\theta})] + \epsilon}\right)^2$, where $\epsilon=10^{-3}$ and $\mathrm{sg}[\cdot]$ indicates a stop-gradient function.
This normalized MSE loss function can be interesting in cases where $\sigma_k^2$ is impractical to obtain. 

The intensity readout at a given pixel may be negative due to the additive Gaussian component in the noise statistics. This can occur in practice in an experiment via background subtraction, when areas on the camera sensor detect only low intensities. As a practical measure to keep the loss function real-valued when calculating the square root of intensity for the Poissonian loss function, negative intensity values are customarily forced to zero~\cite{Cao:17, Kandel2019-hs}. Hence, in both simulation and experimental scenarios, we assign zero to negative intensity values when optimizing $L_{\mathrm{Poisson}}$. However, by zero-cropping the intensity data, we may inadvertently eliminate valuable information, thereby causing a potential bias in our reconstruction results. To examine this bias, and quantify the potential information contained within negative values, we also test $L_{\mathrm{Mixed}}$ with zero-cropped data, as shown in Fig.~\ref{fig:simulation}. The quality of reconstruction in these conditions falls between the results from optimizing $L_{\mathrm{Poisson}}$ with zero-cropped data and $L_{\mathrm{Mixed}}$ with unaltered data. This observation suggests that the enhanced reconstruction quality derived from using a mixed-statistics loss function can partially be credited to the statistical information encapsulated in the negative pixel values resulting from background subtraction.

The results presented in this study underscore the importance of considering mixed Poisson-Gaussian noise statistics in ptychographic image reconstruction. We have demonstrated, through both experimental and simulated data, that using an MLE loss function that considers this mix of noise statistics improves reconstruction quality, particularly in low signal-to-noise ratio conditions. This enhanced performance indicates that the mixed-statistics loss function can extract more information from the measured data by accurately accounting for the underlying noise statistics. An interesting outcome of this study concerns the practice of zero-cropping negative intensity values. We find that this practice introduces a bias into the reconstructions, highlighting the importance of preserving all statistical information in the data. 

It is worth noting that some types of detectors bypass the issue of significant Gaussian readout noise, such as high-performance photon-counting hybrid pixel detectors notably used in x-ray ptychography~\cite{Guizar-Sicairos2014-ze, Pfeiffer2017-vz}. In such cases, optimizing $L_{\mathrm{Poisson}}(\bm{\theta})$ can yield excellent reconstruction results. Future research could focus on studying the convergence behavior of a mixed-statistics loss function when the illumination field is jointly optimized, as we have observed that optimizing $L_{\mathrm{Mixed}}(\bm{\theta})$ occasionally leads to a less reliable convergence when dealing with a poor initial estimate for the illumination field.
To ensure valid comparison and to attain image retrieval under extremely ill-posed conditions, we included the additional step of pre-calibrating the illumination field in this study. Such a step is typically unnecessary in well-posed ptychographic reconstructions~\cite{Maiden2009-pn} or other approaches to noise-robust phase retrieval methods~\cite{Shen2017-zq, Chang2023-xn}.

In summary, the findings presented here could potentially propel significant advancements in the field of computational imaging, leading to improved image retrieval under challenging noise conditions. By offering a more accurate reflection of real-world ptychographic measurements, a loss function that considers mixed Poisson-Gaussian noise statistics could greatly contribute to various fields, including material science, biology, and nanotechnology, where high-quality image reconstruction under low-SNR conditions is critical. Moreover, the utility of using a mixed-statistics loss function is not just limited to ptychography but extends to many computational and gradient-based imaging methods, broadening its applicability~\cite{Antipa2018-ja, OToole2018-qd, Raskar, Hong2004-av, Liutkus2014-vw}.

\textbf{Funding:} Netherlands Organization for Scientific Research NWO (Perspective P16-08).

\textbf{Disclosures:} The authors declare no conflicts of interest.

\textbf{Data availability:} Raw data, source code, and reconstruction scripts are available in \cite{Seifert2023-ak}.

\textbf{Supplemental Document:}
See Supplement 1 for supporting content.


\bibliography{bibliography}

\bigskip



\onecolumngrid
\clearpage
\beginsupplement
\begin{center}
\textbf{\LARGE Maximum-likelihood estimation in ptychography in the presence of Poisson-Gaussian noise statistics: supplemental document}

\bigskip
Jacob Seifert,$^1$, Yifeng Shao,$^{1,2}$ Rens van Dam,$^1$ Dorian Bouchet,$^3$ Tristan van Leeuwen,$^{4,5}$ and Allard P. Mosk$^1$\\ \vspace{0.15cm}
\textit{\small $^\mathit{1}$Nanophotonics, Debye Institute for Nanomaterials Science and Centre for Extreme Matter and Emergent Phenomena, Utrecht University, P.O. Box 80000, 3508 TA Utrecht, The Netherlands}\\
\textit{\small $^\mathit{2}$Imaging Physics Department, Applied Science Faculty, Delft University of Technology, The Netherlands}\\
\textit{\small $^\mathit{3}$Univ. Grenoble Alpes, CNRS, LIPhy, 38000 Grenoble, France}\\
\textit{\small $^\mathit{4}$Centrum Wiskunde \& Informatica, Science Park 123, 1098 XG, Amsterdam, The Netherlands}\\
\textit{\small $^\mathit{5}$Mathematical Institute, Utrecht University, Budapestlaan 6, 3584CD, Utrecht, The Netherlands}
\end{center}
\vspace{1cm}

This document provides supplementary material to \textit{Maximum-likelihood estimation in ptychography in the presence of Poisson-Gaussian noise statistics}.

\section{Derivation of the maximum-likelihood estimation loss functions}
\label{chap:MLE_derivation}

In this supplementary chapter, we provide a derivation of the maximum-likelihood estimation (MLE) loss functions used in a ptychography framework based on automatic differentiation.  MLE operates by selecting the set of parameters that maximize the likelihood function, thus ensuring the best fit to the observed data when the noise follows a known statistical probability distribution. 

\subsection{Poissonian noise statistics}

Let us write the parameters of the ptychography model as a vector $\bm{\theta} = [\theta_1, \theta_2, ..., \theta_{\hat{N}}]$ with $\hat{N}$ denoting the total number of free parameters, presented in our case as complex-valued object pixels.
Within a physics-based forward model of ptychography (as detailed in [1]), we can denote the expected intensity value at a certain pixel $k$ as $I_k(\bm{\theta})$.
In essence, $I_k(\bm{\theta})$ is the noise-free predicted intensity given a specific parameter vector $\bm{\theta}$. 
Considering a discrete random variable $Y_k$, which characterizes the intensity measurement on a camera sensor, we assume a Poisson distribution with an expectation value of $I_k(\bm{\theta})$. The probability mass function is thereby given by
\begin{equation} 
    p(Y_k|\bm{\theta}) = \frac{I_k(\bm{\theta})^{Y_k}}{Y_k!}\exp{(-I_k(\bm{\theta}))},~
    Y_{k \in \{1,\,2,\,...,\,N\}}.
\end{equation}
Assuming that the $N$ measurements are statistically independent, we can express the likelihood function as
\begin{equation}
    \mathcal{L}(\bm{\theta}) = \prod_{k}^N p(Y_k|\bm{\theta}).
\end{equation}
For the sake of computational convenience, we employ the log-likelihood function $\ell(\bm{\theta})$, as the natural logarithm preserves order while transforming the product into the following sum:
\begin{equation}
    \ell(\bm{\theta}) = \ln{\mathcal{L}(\bm{\theta})} = \sum_{k}^N \left( Y_k\ln{I_k(\bm{\theta})} - I_k(\bm{\theta}) - \ln{Y_k!} \right).
\end{equation}
As suggested in Chapter 4.1 of [2], we can now find the second-order Taylor expansion in terms of $\sqrt{I_k(\bm{\theta})}$ at the point $\sqrt{I_k(\bm{\theta})} = \sqrt{Y_k}$ as
\begin{equation}
Y_k\ln{I_k(\bm{\theta})} - I_k(\bm{\theta}) \approx -Y_k + Y_k\ln{Y_k} - 2\left( \sqrt{Y_k} - \sqrt{I_k(\bm{\theta})}\right)^2.
\end{equation}
In practice, we aim to minimize the negative log-likelihood function. This transformation leads to equivalent outcomes and enables us to implement the optimization problem using a reconstruction algorithm based on gradient descent minimization and automatic differentiation libraries such as TensorFlow~[3].
Hence, ignoring all constant additive terms and all multiplicative constants, the MLE loss function for Poissonian noise statistics can be written as
\begin{equation}
    L_{\mathrm{Poisson}}(\bm{\theta})=-\ell(\bm{\theta})=\sum_{k=1}^N\left( \sqrt{Y_k}- \sqrt{I_k(\bm{\theta})}\right)^2.
\end{equation}

\subsection{Gaussian noise statistics}

Minimizing the mean squared error is a common approach to optimization problems and is mathematically closely related to using an MLE loss function with the assumption of Gaussian noise statistics. Even though that is not the ideal assumption for a random variable $W_k$ representing an intensity measurement, it becomes practicable for a large number of detected photons or for cases where $W_k$ can be modeled as a sum of a large number of independent, identically distributed variables, regardless of their underlying distributions (central limit theorem). Considering the random variable $W_k$ following a Gaussian distribution with the mean $I_k(\bm{\theta})$ and constant variance $\sigma^2$, we can express the probability density function as
\begin{equation}
    p(W_k|\bm{\theta}) = \frac{1}{\sqrt{2\pi \sigma^2}}\exp\left(-\frac{(W_k - I_k(\bm{\theta}))^2}{2\sigma^2} \right).
\end{equation}
In analogy to the previous section, we can then express the log-likelihood as
\begin{align}
    \ell(\bm{\theta}) &= \ln{\mathcal{L}(\bm{\theta})} = \ln\prod_{k}^N p(W_k|\bm{\theta}) \\
    \ell(\bm{\theta}) &= -\frac{N}{2}\ln(2\pi\sigma^2) - \frac{1}{2\sigma^2}\sum_{k=1}^N\left( W_k- I_k(\bm{\theta})\right)^2.
\end{align}
Now, by neglecting the constant additive term and multiplicative constants, it becomes evident that $\ell(\bm{\theta})$ can be maximized by the least squares method. We can write the MLE loss function for Gaussian noise statistics as
\begin{equation}
    L_{\mathrm{Gaussian}}(\bm{\theta})=\sum_{k=1}^N\left( W_k- I_k(\bm{\theta})\right)^2.
\end{equation}

\subsection{Mixed Poisson-Gaussian noise statistics}

Continuing from the previous sections, let us consider a random variable $X_k$ as the intensity measurement on a camera sensor with readout noise. We express $X_k$ as the sum over two random variables $X_k = Y_k + Z_k$, where $Y_k$ follows a Poisson distribution of expectation value $I_k(\bm{\theta})$ (see equation~S1) and $Z_k$ follows a centered Gaussian distribution of variance $\sigma_k^2$. The probability density function for $Z_k$ is given by
\begin{equation}
    p_\mathrm{g}(Z_k) = \frac{1}{\sqrt{2\pi \sigma_k^2}} \exp{\left( - \frac{Z_k^2}{2\sigma_k^2} \right)}.
\end{equation}
The random variable $Y_k$ can be approximated as a Gaussian distribution with mean $I_k(\bm{\theta})$ and variance $I_k(\bm{\theta})$:
\begin{equation}
    p_\mathrm{p}(Y_k|\bm{\theta}) \simeq \frac{1}{\sqrt{2\pi I_k(\bm{\theta})}} \exp{\left[ - \frac{(Y_k-I_k(\bm{\theta}))^2}{2I_k(\bm{\theta})} \right]}.
\end{equation}
We can express the probability density function of $X_k$ as
\begin{align}
    p(X_k|\bm{\theta}) &= \int\limits_{-\infty}^{+\infty}  p_\mathrm{p}(\tau|\bm{\theta})p_\mathrm{g}(X_k-\tau) \, \mathrm{d}\tau \\
    p(X_k|\bm{\theta}) &= \frac{1}{\sqrt{2\pi(I_k(\bm{\theta})+\sigma_k^2)}} \exp{\left[ - \frac{(X_k-I_k(\bm{\theta}))^2}{2(I_k(\bm{\theta})+\sigma_k^2)} \right]}.
\end{align}
In analogy to the case without readout noise above, we can now define the MLE loss function as the negative log-likelihood function:
\begin{align}
    L_\mathrm{Mixed}(\bm{\theta}) &= -\ell (\bm{\theta}) = -\ln{\mathcal{L} (\bm{\theta})} = -\ln\prod_{k}^N p(X_k|\bm{\theta}) \\
    L_\mathrm{Mixed}(\bm{\theta}) &= \sum_{k=1}^N\left( \ln[I_k(\bm{\theta})+\sigma_k^2]+\frac{[X_k-I_k(\bm{\theta})]^2}{I_k(\bm{\theta})+\sigma_k^2}\right).
\end{align}
Here, we have neglected all constant additive terms and all multiplicative constants.  To apply this loss function in an optimization framework, we need to obtain $\sigma_k^2$, the variance on the readout noise for each pixel, from dark measurements.

\section{Experimental setup and method}
\label{supp:setup}
\begin{figure}[ht]
\centering
\includegraphics[width=0.9\linewidth]{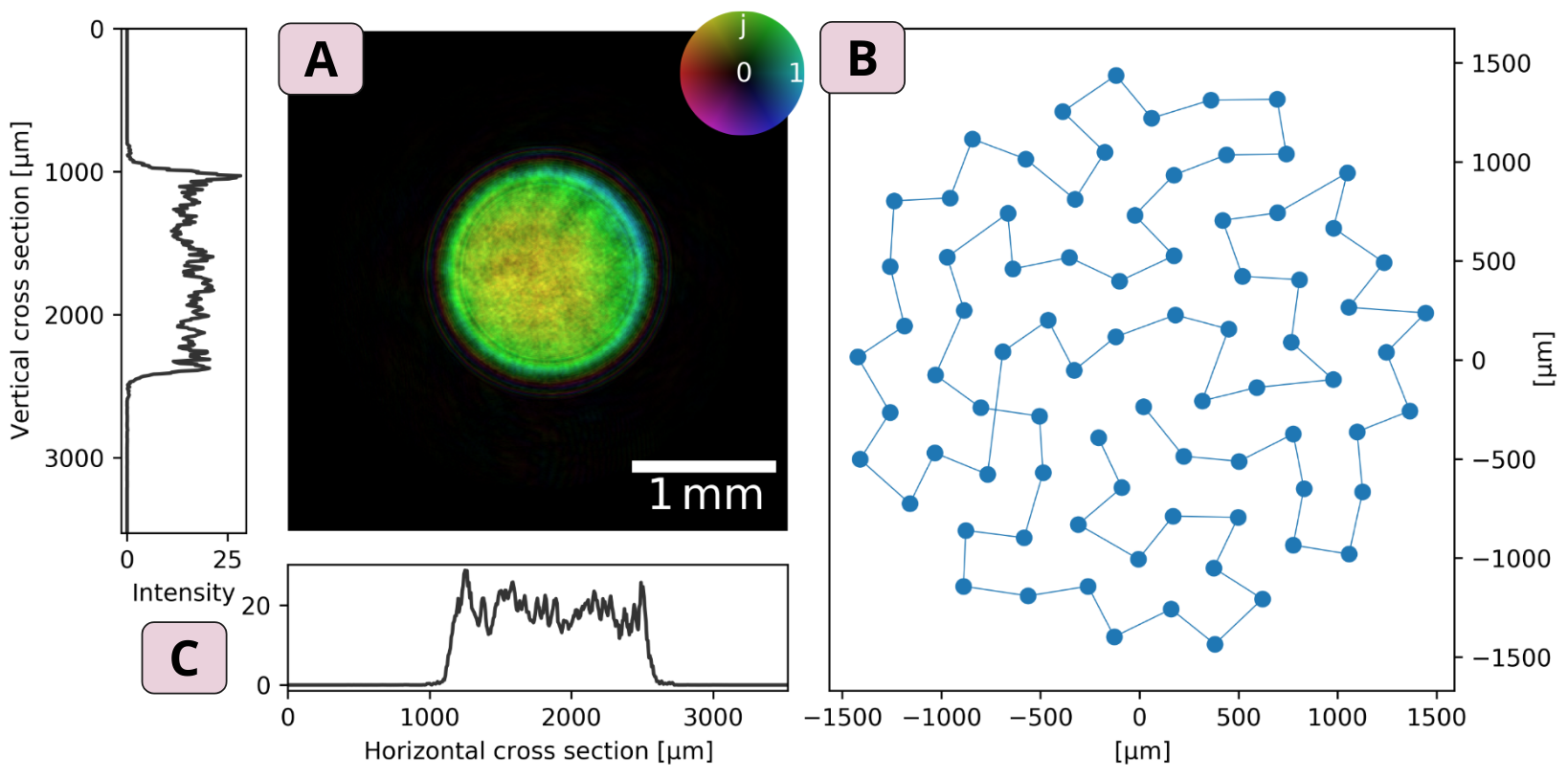}
\caption{(A) Visualization of the complex-valued illumination field that is used in the experiment. The image brightness represents the field amplitude, and the color represents the phase (see circular colorbar). (B) The object's scanning trajectory through the illumination beam in the ptychographic experiment. (C) Horizontal and vertical intensity profiles, centered on the illumination field.}
\label{fig:probe_and_scanningpattern}
\end{figure}

The experimental setup for this ptychography study, also used for numerical simulations, is depicted in Fig~2 of the main document. A coherent laser beam (Cobolt Jive 100\texttrademark) with wavelength $\lambda=\SI{561}{nm}$ is coupled into a single-mode fiber. A fiber collimator (60FC-L-0-M75-26, Schäfter+Kirchoff) expands the beam to around \SI{25}{mm} in diameter, which  then illuminates a \SI{500}{\micro\meter} pinhole.
Using a 2-lens system with a magnification of $M=3$, the pinhole is imaged onto the object, resulting in the illumination field shown in Fig.~\ref{fig:probe_and_scanningpattern}, panel A. 
The two transfer lenses have diameters of \SI{22.9}{mm}, with focal lengths of \SI{5}{mm} and \SI{15}{mm}, respectively.
The object (µChart1951 Test Target, QingYing E\&T LLC) is mounted on a motorized XY-stage with stepper motor actuators (ZFS25B, Thorlabs). The scanning trajectory is shown in panel B of Fig.~\ref{fig:probe_and_scanningpattern}. It comprises a total of 80\,positions in a Fermat spiral pattern to optimize for overlap uniformness~\cite{Huang2014-yo}. Using a traveling salesman algorithm, this trajectory is optimized to minimize total travel distance. The linear overlap between adjacent positions is approximately \SI{60}{\percent}~\cite{Bunk2008-nq}.
The diffraction patterns are recorded \SI{37.7}{mm} behind the object using a CMOS camera (acA2440-35um, Basler) that features a binned pixel size of \SI{6.9}{\micro\meter} and 1024x1024 total pixels. 

To control the Signal-to-Noise Ratio (SNR) in the measurement, we vary the camera exposure time from \SI{30}{\micro\second} to \SI{300}{ms} over 22 steps, covering four orders of magnitude. We derive the spatially varying readout noise variances $\sigma_k^2$ for each exposure time from 300 dark measurements, during which the laser beam is blocked. 
To mitigate Johnson-Nyquist noise fluctuations, we operate the camera sensor in a temperature controlled environment at \SI{21}{\celsius}.
These measurements also provide us with an average background image for each exposure time setting that we subtract from each diffraction pattern. 
To facilitate exact reproduction of the results presented in this study, the raw background and noise statistics data are included alongside the reconstruction framework in~\cite{Seifert2023-ak}.

It is crucial to extract statistical information from potential negative intensity values resulting from Gaussian readout noise. Therefore, we require a black level offset to ensure that no pixel of the sensor ever reads the value of zero in dark measurements. With our Basler camera, we monitor the smallest pixel value for increasing black level settings and observe that an offset of 4 first ensures that all pixel values are larger than zero. To minimize the reduction in dynamic range, we choose this relatively small black level offset for the rest of this work. 

\begin{figure}[ht]
\centering
\includegraphics[width=0.6\linewidth]{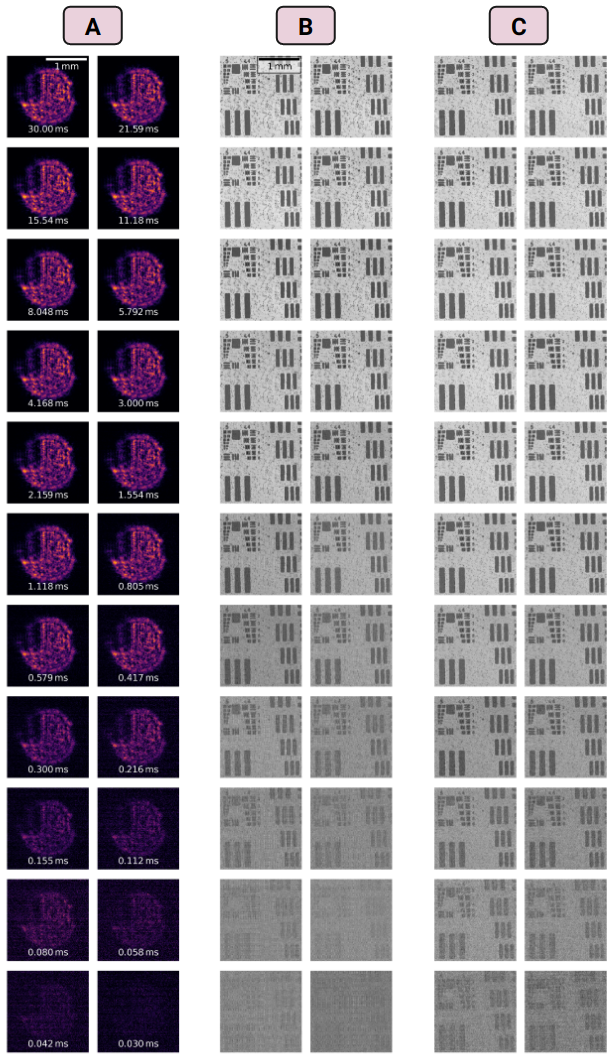}
\caption{Column (A): Visualization of one diffraction pattern from the full data set at varying camera exposure time settings. The exposure times range from \SI{30}{ms} (upper left) to \SI{30}{\micro\second} (lower right). Columns (B and C): Amplitude images reconstructed from the ptychographic data sets with the respective exposure times shown in column A in the same order. Column B is reconstructed with a loss function assuming solely Poissonian noise statistics. Column C is reconstructed with the mixed-statistics loss function assuming Poisson-Gaussian noise.}
\label{fig:full_frame_reconstructions}
\end{figure}

For each scanning point, an independent measurement is obtained for each exposure time setting, and an additional high-SNR measurement is taken by averaging 100 images with the highest exposure time. This high-SNR measurement aids the calibration during the reconstruction phase. In Fig.~\ref{fig:full_frame_reconstructions} presents an expanded version of Fig.~3 from the main manuscript. In the left column, the noise degradation of a single diffraction pattern is shown for all 22 exposure time settings. The central and right columns (B and C) provide a visual comparison of the reconstruction quality for each of these exposure time settings. Specifically, column B showcases reconstructions obtained by using the Poissonian log-likelihood loss function $L_\mathrm{Poisson}(\bm{\theta})$ for optimization, while column C displays reconstructions achieved by employing the mixed Poisson-Gaussian log-likelihood loss function $L_\mathrm{Mixed}(\bm{\theta})$. This comparative illustration provides a clear understanding of the impact of the chosen loss function on the quality of reconstruction across a large range of exposure times.

\section{Reconstruction procedure}

The reconstruction procedure begins with diffraction pattern preprocessing. An experimentally acquired mean dark image is subtracted to correct for background noise and account for hot or dead pixels. In cases where $L_\mathrm{Poisson}$ is used for optimization, negative values are set to zero due to the need for a real-valued loss function. Negative values cannot be incorporated into a noise model that solely assumes Poissonian statistics. 

Typical CCD and CMOS cameras involve an analog-to-digital converter that converts the number of detected photons into analog-to-digital units (ADU). To rectify the assumption that the intensity measurement is Poisson distributed, we rescale the data by the inverse of the overall system gain. In the case of our CMOS camera, the inverse of the overall system gain is specified by the manufacturer to be $\SI{2.7}{\frac{e^{-}}{ADU}}$.

Initially, a high-SNR reconstruction is conducted on the calibration dataset discussed in Section~\ref{supp:setup}. This helps rectify experimental uncertainties such as the object-camera distance and scanning positions, as well as obtaining a high-quality reconstruction of the illumination field (see Fig.~\ref{fig:probe_and_scanningpattern}, panel A). Following this, reconstructions from the lower-SNR diffraction patterns are retrieved using the pre-calibrated illumination field.
Each reconstruction is performed in sequence on a commercial GPU (Nvidia RTX A6000) with the same hyperparameter and regularization settings. Over 100 epochs, the learning rate for the ADAM optimizer~\cite{Kingma2014-fa} starts at $lr=0.1$ and exponentially decays at a rate $\lambda=0.03$, following the schedule $lr_{\mathrm{n}+1} = lr_{\mathrm{n}} \mathrm{e}^{-\lambda}$.

Three regularization terms are added to the loss function, resulting in a final loss function in the form of $L = L_{Poisson/Mixed} + \sum_{i=1}^3L_\mathrm{Reg, i}$. 
\begin{enumerate}
    \item An L1 norm on the amplitudes for the illumination field outside a circular support constraint $\mathcal{S}$ with a radius of \SI{1.5}{mm}.  This regularization term accelerates the convergence of the illumination calibration and is motivated by our experimental setup producing a circular illumination with an approximate radius of \SI{0.75}{mm}. This is expressed as
    \begin{equation}
        L_\mathrm{Reg, 1} = \alpha\sum_{(x, y) \in \mathcal{S}} |P(x, y)|,
    \end{equation}
    where $P(x, y)$ denotes the 2-dimensional illumination field ($P$ for "probe"). The factor $\alpha$ regulates the strength of the regularization, typically chosen as $\alpha \approx 100$ in our calibration procedure.
    
    \item  A minor L1 norm on the amplitudes in the object given by
    \begin{equation}
        L_\mathrm{Reg, 2} = \beta\sum_{(x, y)}^{\hat{N}} |O(x, y)|,
    \end{equation}
    where $O(x, y)$ denotes the 2-dimensional complex-valued object with a total number of pixels $\hat{N}$.
    This regularization drives towards finding a compact solution and mitigates high object amplitudes in the object's boundary areas that are insufficiently illuminated. We set $\beta = 0.0001$.
    
    \item A minor L1 norm on the summed magnitudes of the object in frequency space is expressed as
    \begin{equation}
        L_\mathrm{Reg, 3} = \gamma\sum_{(x, y)}^{\hat{N}} |\hat{O}(u, v)|,
    \end{equation}
    where $\hat{O}(u, v)$ denotes the Fourier transform $\hat{O}(u, v) = \mathcal{FT}\{O(x, y)\}$. We observe that this regularization term can stabilize the optimization using $L_\mathrm{Mixed}$, which sometimes exhibits a poorer convergence behavior than optimizing $L_\mathrm{Poisson}$ or helps prevent numerical divergence phenomena for less-than-optimal learning rates. We set $\gamma = 0.001$.
\end{enumerate}
Note that choosing the regularization prefactors $\alpha, \beta, \gamma$ is arbitrary and depends on non-physical parameters such as the sampling. Therefore, we adopt a heuristical approach to set them small enough to ensure that the data fidelity term strongly dominates the reconstruction process. By doing so, we preserve the valuable comparative basis between the two maximum likelihood estimation (MLE) loss functions while subtly enhancing the reconstructions:
\begin{equation}
    \frac{L_\mathrm{Poisson/Mixed}}{\sum_{i=1}^3L_\mathrm{Reg, i}} \geq 100.
\end{equation}

\end{document}